\documentstyle[prb,aps,graphicx]{revtex}
\begin{document}
\twocolumn
\title{Simulations of the Static Friction Due to Adsorbed Molecules}
\author{Gang He and Mark O. Robbins}
\address{Department of Physics and Astronomy, Johns Hopkins University,
Baltimore  Maryland 21218}
\date{\today}
\maketitle
\begin{abstract}
The static friction between crystalline surfaces separated by
a molecularly thin layer of adsorbed molecules is calculated using
molecular dynamics simulations.
These molecules naturally lead to a finite static friction
that is consistent with macroscopic friction laws.
Crystalline alignment, sliding direction,
and the number of adsorbed molecules are not controlled in most
experiments and are shown to have little effect on the friction.
Temperature, molecular geometry and interaction potentials
can have larger effects on friction.
The observed trends in friction can be understood in terms of a simple
hard sphere model.
\end{abstract}
\narrowtext

\section{Introduction}

The last decade has seen great advances in techniques for
measuring friction in contacts whose geometry and/or chemistry
are controlled with atomic
precision.\cite{carpick97,krim96,granick92,gee90,mcfadden98,bhushan95b}
At the same time, increases in computational power have allowed
increasingly realistic simulations of
friction.\cite{robbins00c,harrison99}
These studies reveal unexpected behavior that raises questions
about the molecular origins of
static friction and the
``laws'' of friction that describe most macroscopic objects.
 
Friction is the lateral force needed to slide one object over another.
The force needed to initiate sliding is called the static friction $F_{\rm s}$.
Its existence
implies that the surfaces have locked into
a local free energy minimum, and $F_{\rm s}$ represents the lateral
force needed to displace them out of this minimum.
The force needed to maintain a constant sliding velocity $v$ is
called the kinetic friction $F_k(v)$, and it represents the force
required to replace energy dissipated during sliding.

Macroscopic objects almost always exhibit a finite static friction
and a kinetic friction that is slightly smaller at low velocities.
One puzzling result from many molecular scale theories of friction between
bare surfaces is that the
static friction almost always vanishes, and is not closely related
to the kinetic friction.\cite{robbins00c,robbins96}
This indicates that some important feature is missing from these model
systems, that must be included to make contact with macroscopic
experiments.
Of course real surfaces are generally not bare, but are coated with
a layer of adsorbed molecules, as well as dust, dirt and other debris.
In this paper we explore the influence of a layer of molecules
between two surfaces on friction forces, and show that including such
layers naturally leads to behavior that is consistent with macroscopic
measurements.

Over three hundred years ago Amontons' published two ``laws'' of 
macroscopic friction
that are still taught and used today.\cite{dowson79}
These state that the friction is proportional to the normal load $L$
pushing two surfaces together, and independent of the apparent
geometrical area of these surfaces $A_{\rm app}$.
Subsequent work
\cite{granick92,gee90,briscoe78,briscoe81,briscoe82,singer92a,berman98,demirel98}
is consistent with a more general friction formula
that is based on the observation\cite{bowden86,dieterich96,berthoud98}
that the actual area of molecular contact
between two surfaces $A_{\rm real}$ is much smaller than the
nominal surface area $A_{\rm app}$.
The friction is assumed to be given by $A_{\rm real}$ times
a local shear stress $\tau$.
If $\tau$ rises linearly with the local
contact pressure $P$,
\begin{equation}
\tau =\tau_0 + \alpha P \ \ \ ,
\label{eq:tauyield}
\end{equation}
then since $P=L/A_{\rm real}$
\begin{equation}
F= \tau A_{\rm real} = \tau_0 A_{\rm real} + \alpha L \ \ \ .
\label{eq:friction}
\end{equation}
This expression agrees with Amontons' laws if $A_{\rm real}$
is proportional to $L$ or if $\tau_0 $ is sufficiently small
compared to $\alpha P$.
The former condition applies if the load is high enough
to produce plastic deformation of the surfaces so that 
$L/A_{\rm real}$ remains equal to the hardness.\cite{bowden86}
It also holds for non-adhesive contacts between ideal elastic
solids with random surface roughness.\cite{greenwood66,volmer97}
However, any adhesive bonding leads to friction in the limit
of zero load and violates Amontons' laws.
For a piece of adhesive tape the first term in Equation \ref{eq:friction}
dominates at low loads, and in many cases friction can be
observed at negative loads.

Early attempts to explain Amontons' laws and the existence
of static friction were based on the
idea that peaks on one surface interlock with valleys
on the other surface.\cite{dowson79,bowden86}
In order to slide,
the top surface must then be lifted up a ramp formed by the
typical slope $\tan \psi $ of the bottom surface.
If there is no microscopic friction between the surfaces, then
the minimum lateral force to initiate sliding is
$F_{\rm s} = L \tan \psi $.
This result satisfies Amontons' laws with a constant coefficient
of friction $\mu_{\rm s} \equiv F_{\rm s}/L = \tan \psi$
that can span all possible values.
However, this geometrical model for friction
can not explain many experimental observations.
For example, coating a single monolayer of surfactant on a surface
does not change its slope, but can reduce the friction
by more than an order of magnitude.
It is also well known that making surfaces too smooth actually
leads to an increase in friction.\cite{bowden86,rabinowicz65}
A practical illustration of this is that
magnetic hard disks are purposely roughened to reduce friction.

The fundamental problem with the above explanation for
the origin of static friction
is that there is no reason for the peaks from
different surfaces to be correlated.
In general one expects that at any instant in time some peaks
will be moving up a ramp and an equal number of peaks will
be moving down.
The net lateral force will average to zero and there will
be no static friction.
A similar problem arises if one imagines that the interlocking
peaks are individual atoms as in the ``cobblestone'' model
discussed by Israelachvili and coworkers.\cite{berman98,israel00}
If atoms are from two identical, aligned surfaces all
of them will go up ramps simultaneously yielding Amontons' laws.
However,
friction measurements are usually made
between misaligned surfaces or surfaces with different lattice
constants.
In this case the force from atoms moving up ramps should cancel
the force from atoms moving downward, yielding a vanishing
static friction.

A number of detailed
analytic\cite{frenkel38,aubry79,bak82,mcclelland89,hirano90}
and simulation
\cite{robbins00c,shinjo93,persson93a,cieplak94,sorensen96,smith96,muser00}
studies have concluded that $F_{\rm s}$ should generally vanish between bare
crystalline or disordered surfaces.
The exception is the case of commensurate surfaces which share
a common periodicity in their plane of contact.
In this case the free energy varies with the phase difference between
the common Fourier components, and simulations of such
surfaces\cite{harrison92b,harrison93b} find a static friction
that rises linearly with pressure as in Eq. \ref{eq:tauyield}.
However the probability that two contacting surfaces are commensurate is
infinitesimally small.
Even identical surfaces are only commensurate when they are perfectly
aligned.\cite{shinjo93}
Any misorientation (Figure \ref{fig:geometry})
causes the surfaces to become incommensurate, i.e.
have no common period.
The free energy is then independent of translations and $F_{\rm s}$ is identically
zero.\cite{aubry79,bak82,hirano90}
If the interactions between the two surfaces are strong enough compared
to interactions within the surfaces
they may deform into a commensurate structure.\cite{aubry79,bak82,muser00}
However, this criterion is unlikely to be met for most of the surfaces around
us, which have been chemically passivated by exposure to air.
Indeed calculations for clean surfaces in vacuum find
zero static friction between different facets of the same
metal!\cite{hirano90,shinjo93,sorensen96,footlast}
Moreover, contacts that are known to be in the elastic limit
exhibit static friction.\cite{granick92,gee90,singer92a,berman98,demirel98}
Edge effects can lead to a finite static friction,
but are not significant for the micron size contacts typically
found between macroscopic objects.\cite{muser01}
Chemical disorder and surface roughness are also unable to
produce observed friction forces.\cite{volmer97,muser01,caroli96,persson96b}

There are relatively few direct experimental measurements of the friction
between bare crystalline surfaces.
However,
studies of adsorbed monolayers,\cite{krim90,krim91}
small crystalline tips,\cite{hirano97} mica,\cite{hirano91}
and molybdenum disulphide\cite{martin93}
support the conclusion that friction vanishes in this limit.
A finite friction has been observed between clean metallic
surfaces,\cite{bowden86,ko00}
but the surfaces were rough and the load appears to have been
high enough to produce plastic deformation.
Further work in this area would be of great value in testing
the above theoretical models.

In recent papers\cite{muser00,muser01,he99,he01}
we have suggested that the so-called ``third bodies''
that are present between most contacting
surfaces\cite{godet84,berthier89}
can provide a general mechanism for static friction.
These third bodies may take the form of wear debris, dust,
or small hydrocarbon or water molecules
that are adsorbed from the air.
The interactions between such bodies are generally weaker
than the interactions within the bounding solids.
This frees them to rearrange at the interface to lower their
free energy and lock the surfaces together.
The layer of third bodies creates an immense number of metastable states,
like that in a glass or granular medium.
The third bodies can always fall into one of these states that is
in registry with both bounding walls and thus produce static
friction.\cite{footpin}

In this paper we report extensive studies of the static friction
produced by layers of spherical or short chain molecules between
crystalline surfaces.
We begin by describing different methods of measuring the static
friction and establishing that it has a well-defined thermodynamic limit.
Then the effect of temperature, wall geometry, interaction
potentials, chain length and areal density
on the static friction are explored.
In each case the static friction obeys Equation \ref{eq:tauyield}
over the experimentally relevant pressure range.
Factors that are not controlled in typical macroscopic experiments
have little influence on $\tau_0$ and even less effect on $\alpha$,
which dominates the friction at high loads.
Such factors include wall geometry, sliding direction, and the length
and density of adsorbed molecules.
Increasing the temperature lowers $\tau_0$,
but has relatively little effect on $\alpha$.
The value of $\alpha$ is strongly dependent on the interaction
potential, particularly the effective hard sphere size of the molecules
compared to that of wall atoms.
All of our results can be understood in terms of
a simple geometrical explanation for the origin of
Equation \ref{eq:tauyield}.

The following section describes the details of our simulations and the
algorithms used to determine $F_{\rm s}$.
Section III determines the effect of each parameter in our
simulations on the static friction
and describes a simple
geometrical picture that explains all of the observed trends.
The final section summarizes the results.

\section{Simulation Method}
\subsection{Potentials and geometry}
\label{sec:potandgeom}

We use a bead spring model\cite{kremer90} that allows us to
explore the behavior of simple spherical molecules or short chains between
bounding solids.
Each molecule contains $n$ spherical monomers.
All monomers interact with each other through a 
truncated Lennard-Jones potential:
\begin{equation}
V(r)=\left\{
\begin{array}{ll}
4 \epsilon [(\sigma/r)^{12}-(\sigma/r)^6
-(\sigma/r_{\rm c})^{12}+(\sigma/r_{\rm c})^6]
&;r<r_{\rm c} \nonumber \\
0&;r>r_{\rm c} \ \ \ \ \ .
\end{array}
\right .
\end{equation}
The parameters $\sigma$ and $\epsilon$
are chosen as the units of length and energy, respectively.
Combined with the monomer mass $m$, they determine the unit of time
$t_{\rm LJ}= (m\sigma^2/\epsilon) ^{1/2}$.
Typical values for hydrocarbons are: $\sigma\sim 0.5$nm, 
$\epsilon \sim 30$meV and $t_{\rm LJ} \sim 3$ps.
We consider the case $r_{\rm c}=2^{1/6}\sigma$ to simulate
a purely repulsive potential,
and $r_{\rm c}=2.2\sigma$ is used to study the effect of adhesive forces.

Monomers are bound into chains by an additional strongly attractive potential 
between nearest neighbors on the same molecule:
\begin{equation}
 V^{\rm CH}(r)=\left\{
\begin{array}{ll}
 -\frac{1}{2}k R_0 {\rm ln} [1-(r/R_0)^2], & r<R_0 \\ \nonumber
 \infty & r\geq R_0 
\end{array}
\right.
\end{equation}
where $R_0=1.5\sigma$ and $k=30 \epsilon \sigma^{-2}$.
These parameters are chosen to prevent chain crossing as described
in previous work.\cite{kremer90}
We consider the case of spherical molecules ($n=1$), and of short
chains with $n=3$ and $6$.
Based on mappings of the bead-spring model to real molecules,\cite{kremer90}
this corresponds to alkane chains with up to roughly 16 carbons.
Substantially
longer molecules are unlikely to have sufficient volatility to contribute
to the airborn contamination that is present on any surface.
Polymers are often placed at interfaces intentionally to act as lubricants.
We have not addressed the behavior of these much longer molecules.

\begin{figure}[tb]
\begin{center}
\leavevmode
\includegraphics[angle=0,width=0.425\textwidth]{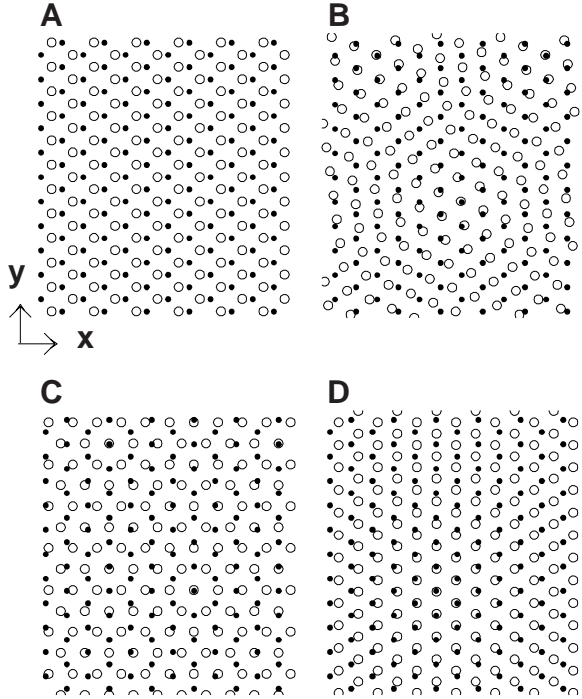}
\vspace{0.5cm}
\caption{
Projections into the $x-y$ plane of atoms from the inner surfaces
of the bottom (solid circles) and top (open circles) walls.
In A through C the two walls have the same structure and
lattice constant, but the top wall has been rotated by $0^\circ$,
$8.2^\circ$, or $90^\circ$, respectively.
In D the walls are aligned,
but the lattice constant of the top wall has been reduced
by 12/13.
The atoms can only achieve perfect registry in the commensurate case A.
The simulation cell was usually at least four times the area shown
here.}
\label{fig:geometry}
\end{center}
\end{figure}

Molecules are confined between two walls formed by the (111)
surfaces of fcc crystals directed normal to the $z$-axis.
Wall atoms are connected to their lattice sites with springs of
stiffness $k_{\rm s}$
in order to model elastic deformation in the simplest way.\cite{thompson90a} 
We consider the completely rigid case, $k_{\rm s}=\infty$, and 
$k_{\rm s}=840\epsilon/\sigma^2$ or $210\epsilon/\sigma^2$.
Both of the latter values correspond to relatively compliant solids.
For example, at $T=0.8\epsilon/k_B$ they lead to root mean squared (rms)
displacements about
lattice sites that are about 4.5 and 9\%, respectively,
of the nearest neighbor separation
(1.2$\sigma$) used in most of our simulations.\cite{lindemannstuff}

Wall atoms and fluid monomers also interact with a Lennard-Jones
potential but with different characteristic energy and length
scales $\sigma_{\rm wf}$ and $\epsilon_{\rm wf}$, respectively.
These are varied to determine the effect of molecular size
and chemistry on static friction.
Direct interactions between atoms on different walls are
not included in the simulations.
They vanish identically for the short
cutoff $r_{\rm c}=2^{1/6}\sigma$ used in most of our simulations,
and have a very small effect on calculated quantities at larger $r_{\rm c}$.
Tests of the effect of wall-wall interactions are discussed briefly in
Section \ref{sec:potentials}.

Periodic boundary conditions are imposed in the plane of the walls.
These periodic boundary conditions prevent us from considering truly
incommensurate systems.
However, the effect of commensurability rapidly decreases as the
length of the common period increases.\cite{robbins00c,muser00,muser01,weiss96}
The degree of commensurability between the walls is varied in two
ways.
The first is to rotate the top wall by an angle $\theta$ about the $z$-axis
(Fig. \ref{fig:geometry} A-C).
Only the range of $\theta$ from 0 to 30$^\circ$ produces inequivalent results.
The second is to retain the alignment of the two lattices ($\theta=0$),
but to change the ratio of the lattice constants of the bottom and top
walls from unity to $12/13$ (Fig. \ref{fig:geometry} D). 

\begin{table}
\caption{Values of the orientation angle $\theta$ and the corresponding
two dimensional basis vectors and number of atoms per surface
layer in the top wall.  Both surfaces have hexagonal symmetry.
The final column gives the percentage
difference between the nearest-neighbor spacings in the two walls.}
\begin{tabular}{|c|c|c|c|c|}
\label{table1}
$\theta$ & $\vec{b_1}/d$ & $\vec{b_2}/d$ & No. of atoms
& $\frac{d-d^\prime}{d}$ ($\%$) \\
\hline
$0^\circ$ & $\vec{i}$ & $\frac{1}{2}\vec{i}+\frac{\sqrt{3}}{2}\vec{j}$&
 $2304$ & 0 \\
\hline
$2.02^\circ$ & $\frac{588}{601}\vec{i}+\frac{12\sqrt{3}}{601}\vec{j}$ &
 $\frac{276}{601}\vec{i}+\frac{300\sqrt{3}}{601}\vec{j}$ &
$2404$ & $-2.1$ \\
\hline
$4.127^\circ$ & $\frac{192}{193}\vec{i}+\frac{8\sqrt{3}}{193}\vec{j}$ &
 $\frac{84}{193}\vec{i}+\frac{100\sqrt{3}}{193}\vec{j}$ &
$2316$ & $-0.3$ \\
\hline
$6.3^\circ$ & $\frac{564}{559}\vec{i}+\frac{36\sqrt{3}}{559}\vec{j}$ &
 $\frac{228}{559}\vec{i}+\frac{300\sqrt{3}}{599}\vec{j}$ &
$2236$ & $1.5$ \\
\hline
$8.2^\circ$ & $\frac{48}{49}\vec{i}+\frac{4\sqrt{3}}{49}\vec{j}$ &
 $\frac{18}{49}\vec{i}+\frac{26\sqrt{3}}{49}\vec{j}$ &
$2352$ & $-1.03$ \\
\hline
$10.4^\circ$ & $\frac{564}{571}\vec{i}+\frac{60\sqrt{3}}{571}\vec{j}$ &
 $\frac{192}{571}\vec{i}+\frac{312\sqrt{3}}{571}\vec{j}$ &
$2284$ & $0.44$ \\
\hline
$14.4^\circ$ & $\frac{564}{589}\vec{i}+\frac{84\sqrt{3}}{589}\vec{j}$ &
 $\frac{156}{589}\vec{i}+\frac{324\sqrt{3}}{589}\vec{j}$ &
$2356$ & $-1.1$ \\
\hline
$19.1^\circ$ & $\frac{20}{21}\vec{i}+\frac{4\sqrt{3}}{21}\vec{j}$ &
 $\frac{4}{21}\vec{i}+\frac{12\sqrt{3}}{21}\vec{j}$ &
$2268$ & $0.79$ \\
\hline
$25.2^\circ$ & $\frac{33}{37}\vec{i}+\frac{9\sqrt{3}}{37}\vec{j}$ &
 $\frac{3}{37}\vec{i}+\frac{21\sqrt{3}}{37}\vec{j}$ &
$2368$ & $-1.4$ \\
\hline
$30^\circ$/$90^\circ$ & $\frac{6}{7}\vec{i}+\frac{2\sqrt{3}}{7}\vec{j}$ &
$\frac{4\sqrt{3}}{7}\vec{j}$ &
$2352$ & $-1.03$ \\
\hline
\end{tabular}
\end{table}

The simulation cell is a rectangle whose height along $y$ is
$\sqrt{3}/2$ times the length along $x$, and
the [110] direction of the bottom wall is directed along the $x$ axis.
Special values of $\theta$ are chosen that allow both
top and bottom surfaces to retain perfect triangular symmetry,
and to have nearly the same nearest-neighbor spacings.
Values for the percentage difference between the spacing on the bottom,
$d$, and on the top, $d^\prime$, walls are given in Table \ref{table1}.
The difference is typically less than $2\%$, and no trends were seen with
the sign or magnitude of this difference.

The system is thermostatted by coupling to a heat reservoir through
Langevin noise and damping terms in the equation of motion.\cite{grest86}
Previous work shows that coupling along the direction of sliding
produces a direct effect on the kinetic
friction, while coupling to the perpendicular
components does not.\cite{robbins00c,smith96,liebsch99}
Thus only the perpendicular components are thermostatted, using the
equation of motion:
\begin{equation}
 m \ddot{\vec{x}}_\perp= \vec{F_{\perp}}-m\Gamma \dot{\vec{x}}_\perp +
 \vec{W}_\perp \ \ \ .
\label{eq:thermo}
\end{equation}
Here $F$ is the force from the interaction potentials,
$\Gamma$ is the friction constant that controls the rate of heat
exchange with the reservoir, and $W$ is the Gaussian-distributed random 
force acting on each monomer.
The rms value of $W$ is determined from $\Gamma$ through the usual
fluctuation-dissipation relation.\cite{grest86}
The equations of motion are integrated using a fifth-order,
Gear predictor-corrector algorithm.\cite{allen87}
In most cases the time step $dt = 0.005t_{\rm LJ}$ and
$\Gamma = 0.4 t_{\rm LJ}^{-1}$.
As discussed in Sec. \ref{sec:thermres},
our measured static friction forces are insensitive to
the choice of $\Gamma$.

The areal density of monomers is specified as the effective coverage
on the separate surfaces before they are brought into contact.
To avoid ambiguities in the definition of the monolayer density,
we define the coverage as the number of adsorbed monomers divided
by the total number of wall atoms in the surface layers of both
walls.
A coverage of 1/2 means that each wall had half as many adsorbed
monomers as surface atoms before the walls were placed in contact.
After contact the monomers from the two walls will produce about
a monolayer between the walls, because the monomers and wall
atoms have roughly the same size in our simulations.

To prepare initial configurations, we typically start with monomers
in a crystalline state, or in the final state of a simulation for
different conditions.
The system is then heated to a temperature of 1.9$\epsilon/k_B$
at a constant pressure of 4$\epsilon \sigma^{-3}$ until the configuration
has randomized (typically 500$t_{\rm LJ}$).
Finally, the temperature and pressure are ramped to the
desired value over 50 $t_{\rm LJ}$ and allowed to equilibrate there
for 450$t_{\rm LJ}$.
Results for the static friction are not sensitive to changes
in this procedure.

\subsection{Determining the static friction}

As noted in the introduction, static friction arises because the system
has managed to lock into a local free energy minimum.
The total energy needed to activate the system out of this free energy
minimum increases with increasing system size.\cite{f3}
However, for any finite system, thermal fluctuations will eventually lead
to activated diffusion in the absence of any lateral force.
This diffusion has been studied previously for the same model
system used here.\cite{muser00}
As discussed by these authors,
the static friction is only well-defined if the limit of
infinite system size is taken before the limit of long times.
If the limits are taken in this order, then the static friction
corresponds to the maximum derivative of the free energy as
the system is driven out of its local minimum.\cite{f3}
If the order is reversed, the static friction is always zero.

Note that even in the infinite-size limit there may be rate
dependence in the measured static friction.
This is because the effective free energy surface depends
on the degree to which monomers can diffuse between the walls.
Based on our recent studies of kinetic friction, we expect 
that monomer diffusion will lead to a weak logarithmic dependence of
static friction on measurement time.\cite{he01}
This is analogous to the logarithmic rate dependence in the apparent
yield stress of glassy systems.\cite{ward83}

Two methods are used to find the static friction. We call them ``ramp''
and ``search'' respectively. In both cases, simulations are done with the
bottom wall fixed and the top wall under constant normal pressure $P$.
A lateral or shear stress $\tau$ is then applied to the top
wall at an angle $\phi$ relative to the $x$-axis.
In the ramp algorithm, the shear stress is increased from zero at a 
constant rate until the wall begins to slide.
The stress at which motion initiates is recorded, and the stress is
then decreased at the same slow rate.
The wall stops moving as the force decreases towards zero, and then begins
to slide again in the other direction when the force is sufficiently
negative.
The magnitude of this threshold force is recorded, and the stress is
increased once more.
This oscillatory process is repeated at least 15 times
to get a statistical sample of threshold forces.
The stress is typically changed at a rate of $0.002\epsilon /\sigma t_{\rm LJ}$.
This rate must be slow enough that the system has time to repin as
the magnitude of the force drops.
It also sets the accuracy on determining the depinning force, because
a finite time is needed before motion of the top wall can be detected.
In most cases the force is increased until the wall reaches a relatively
high sliding velocity ($0.5\sigma/t_{\rm LJ}$) and
the depinning point is identified with the end of the
last previous interval of 20$t_{\rm LJ}$
during which the wall moved less than 0.5$\sigma$.

In the search algorithm, we first guess upper and lower bounds for
the static friction.
Then a lateral stress equal to the average of these values is applied to
determine whether sliding occurs.\cite{f2}
The top wall is considered to be stuck if it moves by less than 1$\sigma$ in
$100 t_{\rm LJ}$ and to be sliding if it moves by more than 5$\sigma$.
If it moves an intermediate distance, the motion is followed
for another $100 t_{\rm LJ}$, and the wall is considered to be sliding if it
moves by more than it did during the first $100 t_{\rm LJ}$.
If the trial stress does not lead to sliding
it becomes the new lower bound,
and if sliding does occur the trial stress becomes the new upper bound.
The system is then returned to an identical initial condition,
and a stress midway between the two new bounds is applied.
This process is repeated enough times (typically five) to obtain
the threshold stress with a precision of better than
$\pm 0.02\epsilon \sigma^{-3}$.
To get good statistics, different initial configurations are
tried.
These are generated by allowing sliding to occur for a distance
of at least 5$\sigma$ and then equilibrating the system at zero lateral
stress for at least 100$t_{\rm LJ}$.

\section{Results}

We first consider how technical details of the simulations may influence
the results.
These include the effects of thermostatting, system size, and the
choice between ramp and search algorithms.
Then the relevant physical parameter space is explored.
First the variations with pressure, temperature and wall stiffness
are discussed.
Then the effects of parameters that are not controlled in
experiments are considered.
These include the wall orientation $\theta$, sliding direction $\phi$,
coverage, and chain length $n$.
Finally, the effect of the potential parameters $\sigma_{\rm wf}$,
$\epsilon_{\rm wf}$,
and $r_{\rm c}$ is explored.

\subsection{Effect of thermostat}
\label{sec:thermres}

As described in the previous section, the wall atoms and monomers are
coupled to a thermostat in order to maintain a constant temperature.
The rate of heat transfer $\Gamma$ is chosen to be small enough that
the thermostat has little effect on the dynamics over the time scale
required for the velocity autocorrelation function to decay to zero.

The main function of the thermostat is to remove heat generated by
sliding friction.
Previous work shows that the thermostat has little effect on kinetic
friction if it is only coupled to the velocity components
that are perpendicular to the sliding
direction.\cite{smith96,liebsch99}
Our focus here is the static friction.
This should be even less sensitive to the thermostat since no
heat is generated until sliding starts, and this is at a force greater
than the static friction.

\begin{figure}[tb]
\begin{center}
\leavevmode
\includegraphics[angle=0,width=0.425\textwidth]{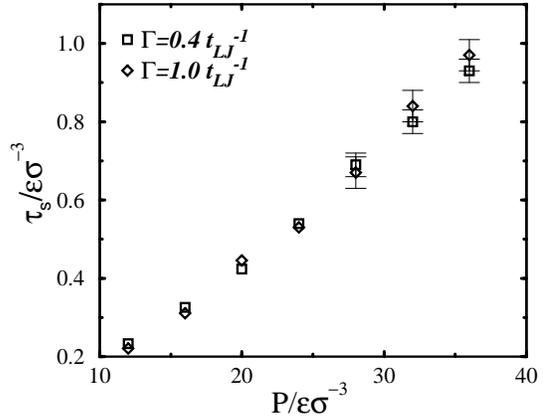}
\caption {Yield stress vs. pressure for two values of the parameter
$\Gamma$ that controls the rate of heat exchange with the thermostat
(Equation \ref{eq:thermo}).
We observed no influence of the thermostat on the yield stress.
For this data, the coverage is 1/2, $\theta=90^\circ$,
$\sigma_{\rm wf}=1.0\sigma$, $\epsilon_{\rm wf}=1.0\epsilon$,
$r_{\rm c}=2^{1/6}\sigma$, $d_{\rm nn}=1.2\sigma$, $T=0.7 \epsilon/k_B$
and $n$=6.
The search algorithm was used, and the simulation cell was 28.8$\sigma$
by 24.9$\sigma$.
Error bars are indicated when they are bigger than the symbol size.
}
\label{fig:f90thermo_rate}
\end{center}
\end{figure}

Fig.$~\ref{fig:f90thermo_rate}$ compares results for the static
friction as a function of pressure $P$ for $\Gamma = 0.4$ and 1.0$t_{\rm LJ}^{-1}$.
The coverage was 1/2 and
the other parameters for these simulations were
$\theta=90^\circ$, $\epsilon_{\rm wf}=\epsilon$,
$\sigma_{\rm wf}=\sigma$, $r_{\rm c}=2^{1/6}\sigma$, and $T=0.7\epsilon/k_B$.
Note that results for the two thermostatting rates are equal within
our statistical error bars.
Other tests show that equivalent results are obtained when only the wall
atoms are coupled to the thermostat.

\subsection{Effect of system size}

The phenomenological expressions for friction described in
the introduction assume that the static friction is given
by a yield stress times the area of contact.
The regions of direct molecular contact between macroscopic solids
typically have diameters of order
microns.\cite{bowden86,dieterich96,berthoud98,rabinowicz65}
It is important to establish that yield stresses calculated at the
much smaller scales of our simulations are representative of larger contacts.
It is also interesting to consider how friction forces would change
if the contact dimensions were reduced to nanometer scales.

\begin{figure}[tb]
\begin{center}
\leavevmode
\includegraphics[angle=0,width=0.425\textwidth]{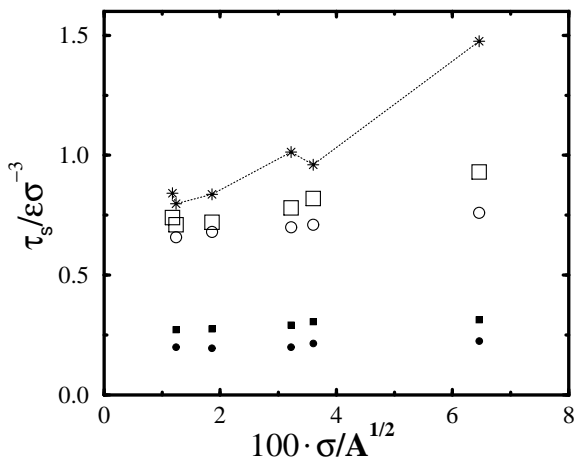}
\caption {Mean yield stress as a function of the inverse square root
of the area $A$ of the periodic simulation cell for the
ramp (squares) and search (circles) algorithms at 
$P=12\epsilon \sigma^{-3}$ (closed symbols)
and $P=28\epsilon \sigma^{-3}$ (open symbols).
The maximum stress needed to initiate sliding from any initial
condition at $P=28\epsilon \sigma^{-3}$ is indicated by asterisks
for the ramp algorithm.
Simulations were done with rigid walls
at $1/2$ coverage, $\theta=90^\circ$,
$\sigma_{\rm wf}=\sigma$, $\epsilon_{\rm wf}=\epsilon$,
$r_{\rm c}=2^{1/6}\sigma$, $d_{\rm nn}=1.2\sigma$, $T=0.7 \epsilon/k_B$ and $n=6$.
Statistical errors are indicated by the symbol size.
}
\label{fig:size}
\end{center}
\end{figure}

Figure \ref{fig:size} compares results from the ramp and search
algorithms and shows how both scale with contact size.
Values of the yield stress $\tau_{\rm s}$ at $P=12$ and $28\epsilon \sigma^{-3}$
are plotted as a function of the inverse width of the simulation cell.
The latter is expressed as $A^{-1/2}$ where $A$ is the area in the x-y plane.
Given $\sigma \sim 0.5$nm, the largest contact widths (smallest values
of $A^{-1/2}$) correspond to 50nm and the smallest widths to about 10nm.

We first consider how contact area affects the variations in
the stress needed to initiate sliding from different initial
configurations.
Points indicated by stars and squares indicate the maximum and
mean forces needed to initiate
sliding for $P=28 \epsilon \sigma^{-3}$ with the ramp algorithm.
Note that the difference between the extreme and average yield
stresses decreases rapidly with system size and is about
10\% for the largest systems.
This indicates that the distribution of yield stresses is well-behaved
and can be thought of as arising from an incoherent sum of contributions
from different regions of the contact.
The same trend is seen for the maximum and mean values from the
search algorithm, but the fluctuations are slightly smaller.

In both algorithms the smallest system sizes tend to produce
larger mean stresses.
However, as assumed in phenomenological theories of
friction,\cite{briscoe78,berman98,bowden86}
the mean values of the yield stress for all pressures and algorithms
go to well defined limits as the contact size diverges.
As shown in Fig. \ref{fig:size},
finite size effects are larger for higher pressures.
This may be because the energy barrier for monomers to move between
metastable states increases with pressure, making them more
likely to get stuck in high energy states.
Finite-size effects also grow rapidly as the coverage decreases,
since there are then fewer monomers at the same wall area.

Note that yield stresses from the ramp algorithm tend to be consistently
higher than those from the search algorithm.
At $P=28\epsilon \sigma^{-3}$ the difference appears to reflect larger
finite-size effects for the ramp algorithm.
However the difference between the two algorithms is nearly independent
of system size for $P=12\epsilon \sigma^{-3}$.
The reason for this discrepancy is that it takes time for the wall to
accelerate once the static friction is exceeded.
One can only be sure that the wall has started sliding when it has moved
by a few $\sigma$.
With the ramp algorithm the force has increased to a higher value by this
point.
Attempts to extrapolate back to the onset of motion only partially corrected
for this offset.
Decreasing the ramp rate reduces the error, but also lengthens the simulation
time proportionately.
The search algorithm reduces the effective ramp rate to zero and efficiently
converges on the yield stress.

For the above reasons the search algorithm is used for most of
the following plots.
Except where noted, the second to smallest system size of Figure
\ref{fig:size} is used.
The surface of each wall then has 576 atoms and the simulation cell
is 28.8$\sigma$ by 24.94$\sigma$ for the usual case of $d_{\rm nn}=1.2\sigma$.
Larger system sizes are used mainly at low coverage in order to
ensure that the deviation of the yield stress from the
thermodynamic limit is comparable to statistical uncertainties.
Larger systems also allow a greater number of rotation angles,
$\theta$, to be studied.

\subsection{Effect of temperature and pressure}
\label{sec:temp}

\begin{figure}[tb]
\begin{center}
\leavevmode
\includegraphics[angle=0,width=0.425\textwidth]{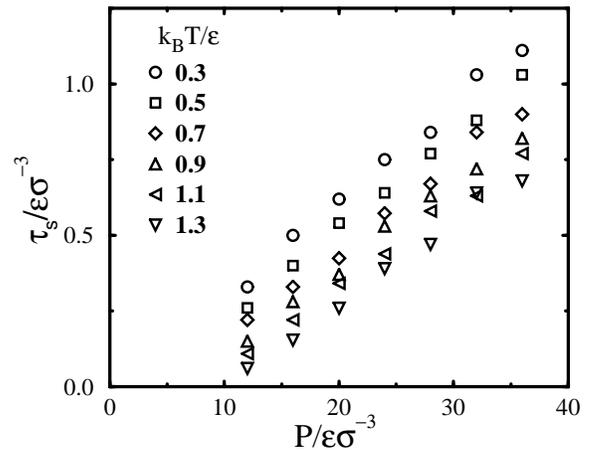}
\caption{Variation of yield stress with pressure at the indicated
temperatures.
In each case, $\tau_{\rm s}$ rises linearly with $P$.
The size of typical statistical errors is indicated by the symbol size.
Errors are slightly larger at the highest pressures.
Simulations were done with the search algorithm for
rigid walls, $1/2$ coverage, $\theta=90^\circ$,
$\sigma_{\rm wf}=\sigma$, $\epsilon_{\rm wf}=\epsilon$,
$r_{\rm c}=2^{1/6}\sigma$, $d_{\rm nn}=1.2\sigma$,
and $n$=6.
}
\label{fig:f90temp}
\end{center}
\end{figure}

Figure \ref{fig:f90temp} shows the yield stress vs. pressure for a
coverage of 1/2 over
a broad temperature range from $k_B T/\epsilon =0.3$ to $1.3$. 
For reference, the triple point and critical temperature of Lennard-Jones
monomers with $r_{\rm c} \rightarrow \infty$
are about 0.7 and 1.3 $\epsilon/k_B$,
respectively,\cite{allen87,smit92}
and the zero pressure glass transition temperature for short Lennard-Jones
chains is between 0.4 and
0.5$\epsilon/k_B$.\cite{baljon97,robbins00a,bennemann98}

Note that $\tau_{\rm s}$ increases linearly with pressure at each $T$,
as assumed in Equation \ref{eq:tauyield}.
For representative values of $\epsilon$=30meV and $\sigma=0.5$nm,
the unit of pressure $\epsilon \sigma^{-3} \sim $40MPa.
Thus the highest pressures in Figure \ref{fig:f90temp}
correspond to about 1.5GPa.
In those cases that we checked, linearity extended up to
much larger pressures (at least 100$\epsilon \sigma^{-3}$).
However, these pressures were not routinely studied because a smaller time
step is needed to integrate the equations of motion at larger pressures.

\begin{figure}[tb]
\begin{center}
\leavevmode
\includegraphics[angle=0,width=0.425\textwidth]{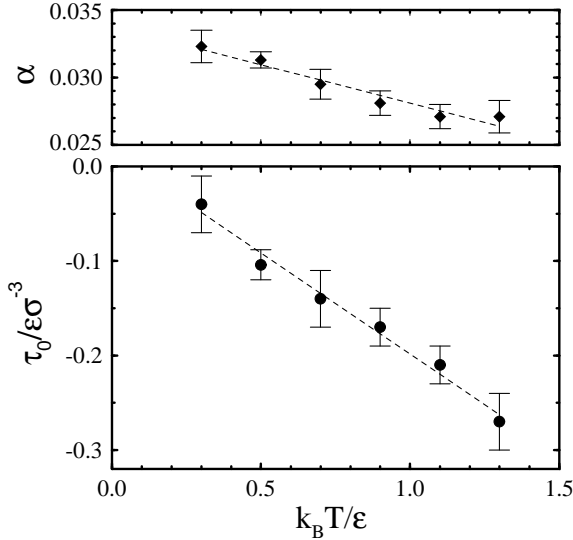}
\caption {Values of $\alpha$ and $\tau_0$ as a function
of temperature from fits of Equation
\ref{eq:tauyield} to the data of Figure \ref{fig:f90temp}.
Dashed lines show the linear fits to the temperature dependence
given in Equation \ref{eq:alphat}.
}
\label{fig:f90alphatau_T}
\end{center}
\end{figure}

The plots of yield stress vs. pressure at different temperatures in
Figure \ref{fig:f90temp} are nearly parallel, but their intercepts shift
rapidly with temperature.
Results from fits to Equation \ref{eq:tauyield} are shown in Figure
\ref{fig:f90alphatau_T}.
The slope $\alpha$ drops by only 16\% with increasing $T$,
while the intercept $\tau_0$ is roughly proportional to $T$.
Linear fits as a function of temperature give:
\begin{equation}
\begin{array}{lll}
	\alpha &=&(-0.0057\pm 0.0007) \frac{k_B T}{\epsilon} 
		+(0.0338\pm 0.0006) \nonumber \\
	\frac{\tau_0}{(\epsilon \sigma^{-3})}&=
		&(-0.214\pm 0.013) \frac{k_BT}{\epsilon} +(0.015\pm0.011).
\label{eq:alphat}
\end{array}
\end{equation}
An explanation for the nearly constant value of $\alpha$ is presented
in Section \ref{sec:potentials}.
In the remainder of this subsection we focus on the variation of $\tau_0$.

At each $T$,
the linear fit for yield stress vs. pressure crosses through zero
at a threshold pressure $P_{\rm t}=-\tau_0/\alpha$.
Above this pressure we find a well defined yield stress in the thermodynamic
limit.
Below $P_{\rm t}$, the adsorbed layer acts like a lubricating liquid,
with a shear stress that vanishes as the shear rate decreases.\cite{he01}
In this regime, the bounding walls are far enough apart that molecules
diffuse freely and do not lock the two walls in a local energy minimum.
From the fit parameters in Fig. \ref{eq:alphat} we see that
$P_{\rm t}$ rises roughly linearly with $T$.
This is because the simulations used $r_{\rm c}=2^{1/6}\sigma$, which
corresponds to the hard sphere limit.
In this regime pressures are balanced only by entropic forces
and thus must scale linearly with temperature.
Attractive interactions are considered in Sec. \ref{sec:potentials}

Most of the following simulations were done at $T=0.7\epsilon/k_B$.
Given the results just described, we expect that the same trends would
be observed at other temperatures.

\subsection{Effect of wall stiffness }
 
As described in Section \ref{sec:potandgeom},
wall atoms were connected to lattice sites by springs
with stiffness $k_{\rm s}$, or fixed rigidly to
these sites ($k_{\rm s} \rightarrow \infty$). 
Decreasing $k_{\rm s}$
has two competing effects on the static friction.
One is to make it easier for the two surfaces to deform into a structure
with a common period.
This tends to increase the yield stress.
However, studies with bare walls found that unrealistically small
values of $k_{\rm s}$ were needed to produce a finite static friction.\cite{muser00}
Decreasing $k_{\rm s}$ also decreases the yield stress by allowing wall atoms
to be pushed out of the way.
In our simulations the second effect always dominates, and $\tau_{\rm s}$
decreases as $k_{\rm s}$ decreases.
This is consistent with a recent analytic theory for static
friction.\cite{muser01}

\begin{figure}[tb]
\begin{center}
\leavevmode
\includegraphics[angle=0,width=0.325\textwidth]{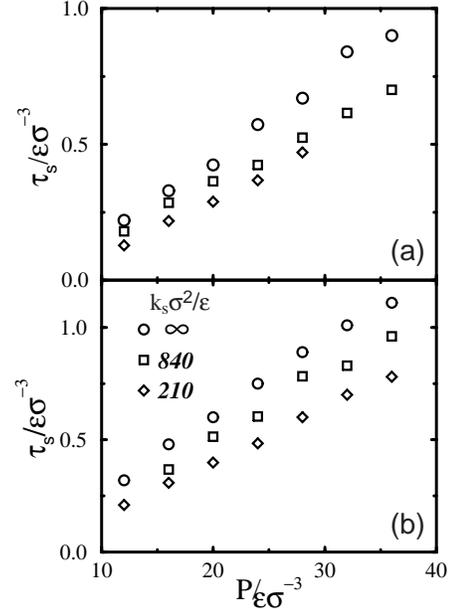}
\caption {Yield stress vs. pressure for the indicated values of the
stiffness $k_{\rm s}$ of the
springs coupling wall atoms to lattice sites.
The top panel shows results for $\theta=90^\circ$ (Fig. \ref{fig:geometry}C)
and the bottom panel shows results for $\theta=0^\circ$ but
$d/d'=12/13$ (Fig. \ref{fig:geometry}D).
In each case,
decreasing the stiffness decreases the yield stress.
These results are from the search algorithm with
$1/2$ coverage, $\sigma_{\rm wf}=\sigma$, $\epsilon_{\rm wf}=\epsilon$,
$r_{\rm c}=2^{1/6}\sigma$, $d_{\rm nn}=1.2\sigma$, and $n$=6.
}
\label{fig:mobility}
\end{center}
\end{figure}

Fig.$~\ref{fig:mobility}$ shows results for systems $C$ and $D$ of
Fig. \ref{fig:geometry} at $k_B T=0.7\epsilon$.
In each case, the yield stress rises linearly with pressure and
decreases monotonically with $k_{\rm s}$.
The value of $\alpha$ drops by up to 50\% as $k_{\rm s}$ decreases from
infinity to $210\epsilon \sigma^{-3}$.
Note that for $k_{\rm s}=210\epsilon \sigma^{-3}$ the rms displacement
about lattice sites is near the usual Lindemann
criterion for melting,
\cite{lindemannstuff}
so much weaker values would be unphysical.
To minimize computer time
and the number of free parameters, we used rigid walls
$k_{\rm s} \rightarrow \infty$ in the following sections.

\subsection{Effect of wall orientation, coverage and chain length }
\label{sec:cov}

Calculated values of the static friction for bare surfaces vary
dramatically with wall geometry.\cite{robbins00c,sorensen96,hirano93}
The yield stress equals that of a bulk crystal
for identical aligned surfaces ($\theta=0^\circ$)
and is identically zero for incommensurate orientations.
Most experiments do not control the wall orientation, and yet
measured friction coefficients between nominally dry surfaces
vary by roughly 10\% within the same laboratory and
by 20\% between laboratories.\cite{rabinowicz65,czichos89}
Another factor that is not controlled in most experiments
is the density and chemistry of adsorbed molecules, although
humidity is controlled in some cases.
If adsorbed molecules are crucial in determining the static
friction, the friction they produce must be relatively
insensitive to the wall orientation, sliding direction,
coverage and chain length.

\begin{figure}[htbp]
\begin{center}
\leavevmode
\includegraphics[angle=0,width=0.325\textwidth]{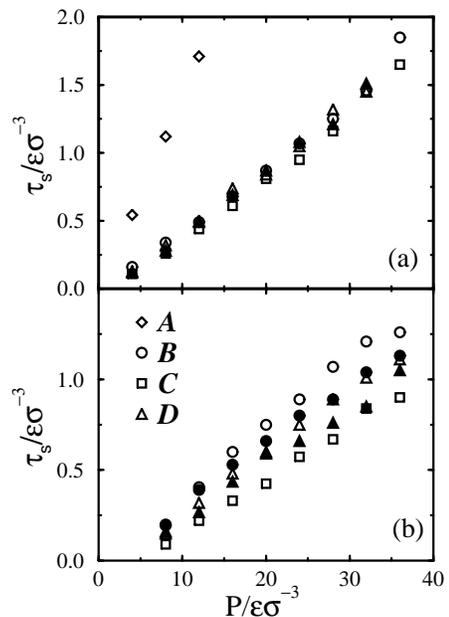}
\caption {Shearing direction and orientation dependence for
(a) 1/8 coverage and (b) 1/2 coverage.
The symbol shape indicates which of the geometries in
Fig. \ref{fig:geometry} were used (see legend).
Open and closed symbols are for lateral force applied along the
$x$ ($\phi=0^\circ$) and $y$ ($\phi=90^\circ$) directions, respectively.
The ramp algorithm was used in (a) and the search algorithm
was used in (b).
The potential parameters were
$\sigma_{\rm wf}=\sigma$, $\epsilon_{\rm wf}=\epsilon$,
$r_{\rm c}=2^{1/6}\sigma$, $d_{\rm nn}=1.2\sigma$, and $n$=6.
}
\label{fig:oridir}
\end{center}
\end{figure}

Figure \ref{fig:oridir} shows the yield stress as a function of
pressure for the different walls shown in Figure \ref{fig:geometry}.
At low coverages (Fig. \ref{fig:oridir}(a)), the friction
is nearly identical for all incommensurate walls.
The friction for commensurate walls is roughly four
times higher, but much less than the value for bare walls.
At a coverage of 1/2 (Fig. \ref{fig:oridir}(b)),
the difference in the yield stresses for different incommensurate
walls increases to roughly 20\%.
A similar variation with the direction of sliding
($\phi=0$ or 90$^\circ$) is seen.
Most of this variation is due to changes in $\tau_0$
and all the curves are nearly parallel.
Fit values of $\alpha$, which dominates the high pressure
friction, only change by about 5\%.

The greater variability at coverage of 1/2 arises because this case
corresponds to a dense monolayer between the surfaces.
Molecules can not pick and choose which portions of the surface to
cover.
As a result the friction is higher for geometries B and D,
where there are larger regions where the walls look commensurate
and molecules can easily lock in registry with both surfaces.
The friction is smallest for geometry C, where there is almost
no region where the two walls are in registry.

\begin{figure}[htbp]
\begin{center}
\leavevmode
\includegraphics[angle=0,width=0.425\textwidth]{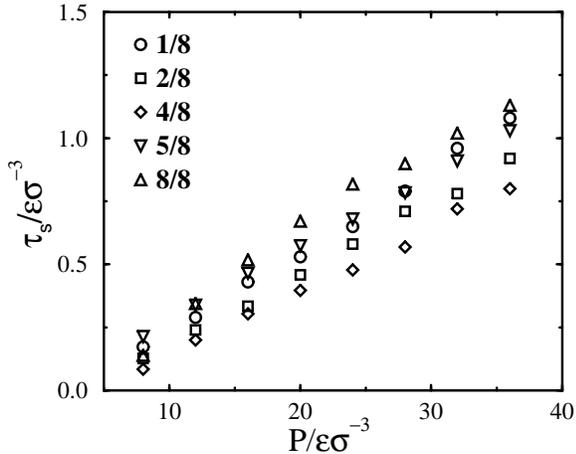}
\caption {Yield stress vs. pressure for the indicated coverages with
system C ($\theta=90^\circ$) where the largest variation was observed.
Note that a coverage of 1/2 on each separated surface produces
about a monolayer when the surfaces are placed in contact.
Results are from the search algorithm with
$\sigma_{\rm wf}=\sigma$, $\epsilon_{\rm wf}=\epsilon$,
$r_{\rm c}=2^{1/6}\sigma$, $d_{\rm nn}=1.2\sigma$, and $n$=6.
}
\label{fig:90coverage}
\end{center}
\end{figure}

Figure \ref{fig:90coverage} shows the yield stress vs. pressure at
different coverages for
the extreme case of geometry C ($\theta=90^\circ$).
Note that the yield stress exhibits the usual linear pressure dependence
at all coverages, and that most of the variation with coverage is in the
intercept $\tau_0$.
The value of $\alpha$ stays within 15\% of the mean value, 0.030.
At each pressure,
the friction decreases as the coverage rises from 1/8 to 1/2.
During this increase in coverage,
more and more of the sites with poor registry must be occupied.
The friction goes down to a minimum value at a complete monolayer
which corresponds to about 1/2 coverage.
If the difference between the sizes of the monomers and wall atoms
was bigger, locking would be frustrated even in the areas with
good registry.
This might lead to even smaller variations with coverage than
those shown here.

As coverage rises from 1/2 to 1 (about two monolayers), the friction
rises once more.
The reason is that two layers of molecules have more internal
degrees of freedom and can lock more easily into simultaneous
registry with both walls.
For this range of coverages yield continues to localize
at the surfaces of the walls.
Previous work on thicker films shows that yield can move into
the film as thickness increases.\cite{thompson95,manias96}
In this limit $\tau_{\rm s}$ approaches the bulk yield stress of
the material making up the adsorbed layer, and eventually becomes
independent of the coverage.
Note that the bulk yield stress of polymeric materials is also
found to follow the linear pressure dependence of Equation \ref{eq:tauyield}. 
\cite{ward83}

\begin{figure}[htbp]
\begin{center}
\leavevmode
\includegraphics[angle=0,width=0.35\textwidth]{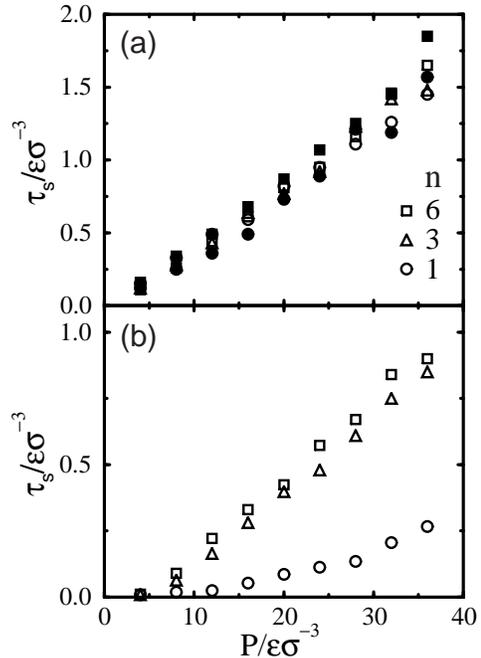}
\caption {Yield stress vs. pressure for the indicated values of
chain length $n$ at coverage of (a) $1/8$ and (b) $1/2$.
Open (closed) symbols are for system C (B).
The potential parameters were
$\sigma_{\rm wf}=\sigma$, $\epsilon_{\rm wf}=\epsilon$,
$r_{\rm c}=2^{1/6}\sigma$, and $d_{\rm nn}=1.2\sigma$.
Errorbars are comparable to the system size and the ramp (search)
algorithm was used for the top (bottom) panel.
Except for spherical molecules at $1/2$ coverage, the friction is
insensitive to chain length.
}
\label{fig:numchn}
\end{center}
\end{figure}

Figure \ref{fig:numchn} illustrates the interplay between the
chain length and coverage dependence.
At small coverages (Fig. \ref{fig:numchn} (a)),
there is little dependence on chain length or wall orientation.
For a complete monolayer (Fig. \ref{fig:numchn} (b)),
chains with $n=3$ and $6$ show nearly the
same behavior, while spherical molecules ($n=1$) show a dramatically
reduced friction and $\alpha$.
This suggests that the spheres are more constrained by their neighbors
from locking into registry with the walls.
This may reflect the fact that although all films contain the same
number of monomers at a coverage of 1/2,
the spheres are closer to their true
equilibrium monolayer density than the chains.
The reason is that the strong bonds within chain molecules lower the
equilibrium spacing and thus increase the equilibrium density at
a given pressure.
The presence of two different types of bonds in chain molecules
may also improve their ability to rearrange to lock into both surfaces.
Another possibility is that chains produce a greater friction because
of the strong intramolecular bonds.
If one monomer in the chain is locked to the bottom wall
and another is in registry with the top wall, the intramolecular bonds
will lock the two walls together.
This process has no analog for monomers and is particularly important
as the coverage increases beyond a monolayer.

\begin{figure}[htbp]
\begin{center}
\leavevmode
\includegraphics[angle=0,width=0.425\textwidth]{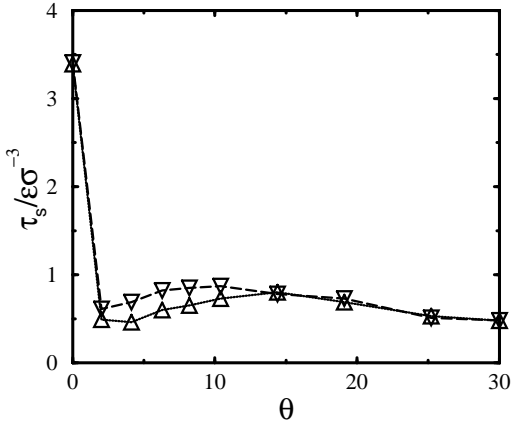}
\caption{Yield stress as a function of orientation angle
$\theta$ at $P=24\epsilon\sigma^{-3}$.
The commensurate case corresponds to $\theta=0$ and the
smallest non-zero angle shown here is $2.02^\circ$.
These results are from the search algorithm with
$1/2$ coverage, $\sigma_{\rm wf}=\sigma$, $\epsilon_{\rm wf}=\epsilon$,
$r_{\rm c}=2^{1/6}\sigma$, $d_{\rm nn}=1.2\sigma$, and $n$=6.
The simulation cell was $57.6\sigma$ by $49.9\sigma$.
}
\label{fig:staticangle}
\end{center}
\end{figure}

Figure \ref{fig:staticangle} shows the static friction as a function of
$\theta$ at a single pressure, $P=24\epsilon \sigma^{-3}$.
This large pressure was chosen to maximize the contribution of
$\alpha$ to the yield stress.
Only $\theta$ between 0 and 30$^\circ$ are shown because other
angles are related by symmetry.

The commensurate case $\theta=0^\circ$ shows a much larger friction
than any of the incommensurate systems.
Even a $2^\circ$ rotation reduces the friction to a value that
is comparable to those at other values of $\theta$.
The same sharp transition was seen for coverages from 1/8 to 1.
The data shown
is for coverage of 1/2 where the variation of the yield stress
at incommensurate $\theta$ has already been shown to be largest.
Even in this case the variation is
less than $\pm 20$\%, which is comparable to experimental variations.
The smaller variations of yield stress with $\theta$ at a coverage of 1/8 are
comparable to our statistical errorbars.

The sliding direction $\phi$ also has relatively little influence
on the yield stress between incommensurate surfaces.
Two orthogonal directions $\phi=0^\circ$ and 90$^\circ$ are shown in
Fig. \ref{fig:staticangle}.
The difference between them is most pronounced at small $\theta$,
where $\phi=0^\circ$ corresponds to pulling nearly parallel
to the nearest-neighbor direction in each wall, and $\phi=90^\circ$
to pulling along the perpendicular direction.

\begin{figure}[htbp]
\begin{center}
\leavevmode
\includegraphics[angle=0,width=0.425\textwidth]{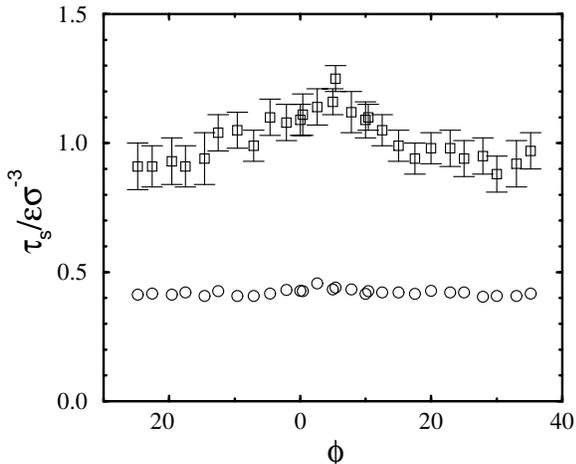}
\caption {The yield stress vs. the angle $\phi$ between the applied stress 
and the $x$ axis for $P=12\epsilon \sigma^{-3}$ (circles)
and 32$\epsilon \sigma^{-3}$ (squares).
These results are from the search algorithm with
$1/2$ coverage, $\sigma_{\rm wf}=\sigma$, $\epsilon_{\rm wf}=\epsilon$,
$r_{\rm c}=2^{1/6}\sigma$, $d_{\rm nn}=1.2\sigma$ and $n=6$.
The simulation cell was $57.6\sigma$ by $49.88\sigma$.
Errorbars are only shown when they are larger than the symbol size.
}
\label{fig:direction}
\end{center}
\end{figure}

Figure \ref{fig:direction} shows the variation with $\phi$ in more
detail for coverage of 1/2 and $\theta=10.4^\circ$,
where the variation in Fig. \ref{fig:staticangle} is large.
The data are symmetric about $\theta/2$ which is the direction
of a 2-fold axis.
This direction gives the maximum friction, and the yield stress
drops to a plateau value on either side.
The variation is within our statistical errors at low pressures
and is less than $\pm 20$\% at the largest pressures studied.
As in Figures \ref{fig:oridir}(b) and \ref{fig:staticangle},
sliding along the x axis
leads to higher friction than sliding along y (equivalent to
$\phi=0$ and $30^\circ$, respectively).

Note that when the yield stress is exceeded the top wall does
not always move exactly in the direction of the applied force.
This effect is most pronounced for the commensurate case,
but is also seen at small $\theta$.

\subsection{Effect of Interaction Potentials}
\label{sec:potentials}

Experimental results for different sliding surfaces and different
surface layers show much larger variations than the fluctuations
for a given material system.
Such changes in material would correspond to changes in the parameters
in our interaction potentials
such as the strength of the interaction between the
wall and fluid $\epsilon_{\rm wf}$, the characteristic
length of this interaction $\sigma_{\rm wf}$, the
range of the interaction $r_{\rm c}$, and
the nearest-neighbor spacing $d_{\rm nn}$ in the walls.
Other structural properties and the elastic moduli and hardness
of the solids
may also be important, but are not easily included within our model.


\begin{figure}[htbp]
\begin{center}
\leavevmode
\includegraphics[angle=0,width=0.35\textwidth]{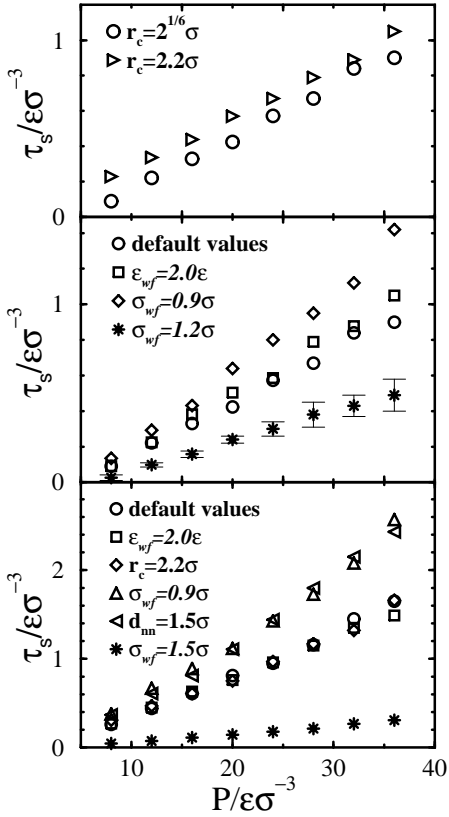}
\caption {Yield stress vs. pressure for
various choices of the potential parameters in system C with $n=6$.
The coverage is $1/2$ in (a) and (b) and $1/8$ in (c).
Unless noted the default values of
$\sigma_{\rm wf}=\sigma$, $\epsilon_{\rm wf}=\epsilon$,
$d_{\rm nn}=1.2\sigma$, and $r_{\rm c}=2^{1/6}\sigma$ are used.
Each of these parameters was changed in turn and the corresponding
results are labeled by the changed value.
The search algorithm was used for the data in (a) and (b),
and the ramp algorithm was used in (c).
}
\label{fig:90potential}
\end{center}
\end{figure}

The results shown so far were all for purely repulsive potentials,
$r_{\rm c}=2^{1/6}\sigma$.
Increasing the cutoff to $r_{\rm c}=2.2\sigma$ includes most of the attractive
tail in the Lennard-Jones potential.
This attractive tail leads to an adhesive force pulling the walls together
in addition to the external pressure $P$.
As shown in Fig. \ref{fig:90potential}(a) and (c),
the yield stress for $r_{\rm c}=2.2\sigma$
is parallel to results for $r_{\rm c}=2^{1/6}\sigma$ and could be obtained
by shifting the zero of pressure to reflect the adhesive stress.
The required shift increases with increasing coverage because the
density of adsorbed molecules that contribute to the adhesive
interaction increases.

When the above results are fit to Eq. \ref{eq:tauyield}, the value
of $\alpha$ is independent of $r_{\rm c}$, and the value of $\tau_0$ increases
with the amount of adhesion.
For purely repulsive interactions $\tau_0$ is negative because a finite
pressure is needed to lock the adsorbed layers in a glassy state
(Sec. \ref{sec:temp}).
When $r_{\rm c}$ is increased to 2.2$\sigma$, a glass forms at
zero pressure, and the value of $\tau_0$ is positive.
Many authors refer to $\tau_0$ as the adhesive contribution to friction,
and it is obvious that $\tau_0/\alpha$
represents the effective
adhesive pressure that must be added to the external pressure to
get strict proportionality between the friction and total load.

These simulations did not include the direct interactions between
atoms from different walls.
As noted in Section \ref{sec:potandgeom}, the wall spacing is
large enough that the direct interaction vanishes for
$r_{\rm c} = 2^{1/6}\sigma$.
Even for $r_{\rm c} = 2.2\sigma$, wall interactions have little effect.
For example the extra adhesive pressure due to wall/monomer interactions
is $\tau_0/\alpha \sim 5 \epsilon \sigma^{-3}$ for 1/2 coverage
(Fig. \ref{fig:90potential}(a)).
Direct calculation of the
extra adhesive pressure due to interactions between atoms on
different walls yields a value that rises from
0.4 to 1.3$\epsilon \sigma^{-3}$ over the
observed range of wall separations.
Correcting the external pressure to account for this interaction
would change $\alpha$ by less than 3\%.

Panels (b) and (c) of Fig. \ref{fig:90potential} show that
doubling the strength of the potential between the
surfaces and monomers has almost no effect on the yield stress.
The main factor that changes $\alpha$ is the ratio between
the characteristic separation between wall and fluid atoms, $\sigma_{\rm wf}$,
and the nearest-neighbor spacing between wall atoms, $d_{\rm nn}$.
The value of $\alpha$ rises monotonically as $\sigma_{\rm wf}/d_{\rm nn}$
decreases.
For the cases shown, $\alpha$ changes by an order of magnitude.
Smaller values of $\alpha$ are obtained by increasing
$\sigma_{\rm wf}/d_{\rm nn}$
further, but it is difficult to increase $\alpha$ for this geometry.
The reason is that the adsorbed molecules begin to penetrate into the
wall when $\sigma_{\rm wf}/d_{\rm nn}$ is too small.
Other crystallographic orientations of the surface, such as (100), may
produce larger values of $\alpha$.
Flat but disordered surfaces produce values of $\alpha$ as high
as 0.3.\cite{muser01,he99}

The strong effect of $\sigma_{\rm wf}/d_{\rm nn}$ and weak effect of
$\epsilon_{\rm wf}/\epsilon$ suggest a geometric origin of $\alpha$
that is analogous to early geometric theories for Amontons' laws
\cite{dowson79}
and to Israelachvili's cobblestone model.\cite{berman98,israel00}
At pressures where the adsorbed layer acts like a glass,
the repulsive part of the Lennard-Jones interaction dominates.
The repulsive force between monomers and wall atoms at a separation $r$
is given by
$F_{\rm rep} = 48 (\epsilon_{\rm wf}/\sigma_{\rm wf}) (\sigma_{\rm wf}/r)^{13}$.
The mean value of $F_{\rm rep}$ must scale with the pressure.
Hence, increasing $P$ by a factor of 2
only decreases the typical separation between monomers and
wall atoms by a factor of $2^{1/13}$ or 5\%.
Increasing $\epsilon_{\rm wf}$ by a factor of two leads to the same
small decrease in separation.
In contrast, the separation changes by a factor of $2^{12/13} \approx 1.9$
with a factor of 2 change in $\sigma_{\rm wf}$.
Thus monomers and wall atoms can be thought of as hard spheres whose
diameter $D$ is nearly independent of pressure and $\epsilon_{\rm wf}$
but directly proportional to $\sigma_{\rm wf}$.

\begin{figure}
\begin{picture}(000,85)
\includegraphics[angle=0,width=0.35\textwidth]{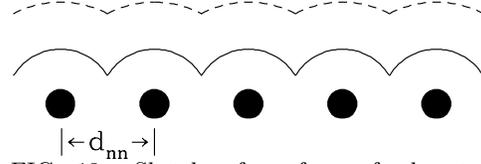}
\end{picture}
\caption {Sketch of surface of closest approach for
$D/d_{\rm nn}=0.7$ (solid line) and 1.3 (dashed line).
The lateral force required to lift monomers up the ramps
that these surfaces represent is given by the slope times
the normal force.
}
\label{fig:sketch}
\end{figure}

Figure \ref{fig:sketch} illustrates the effect of changing the
ratio between the hard sphere diameter and the nearest-neighbor
spacing $d_{\rm nn}$.
The hard-sphere repulsion leads to a surface of closest approach
between a monomer and the wall.
In order for the monomer to slide relative to the wall it must be
lifted up the ramp formed by this surface.
The lateral force required to move monomers up the ramp is just
the normal load times the slope of the surface.
Thus $\alpha$ will be proportional to some average of the slope
over all monomers.
As $D/d_{\rm nn}$ decreases, monomers penetrate more deeply between
the surfaces and both the slope and $\alpha$ increase.
This is precisely the trend seen in Fig. \ref{fig:90potential}. 
Thus the hard-sphere model explains the linear dependence of $\tau$
on pressure as well as the insensitivity to $\epsilon_{\rm wf}$
and the variation with $\sigma_{\rm wf}$ and $d_{\rm nn}$.

\section{Summary and Conclusions}
\label{sec:conc}

The results presented in this paper explore the effect of a
specific type of third-body, small adsorbed molecules, 
on the friction between the bounding solids.
The effect of temperature, coverage, crystalline alignment,
lattice constant, system size, molecular size and
interaction potential parameters were all examined.

The first important result is that static friction is observed
for all systems -- even when the static friction between bare walls
is identically zero.
The shear stress needed to initiate sliding approaches a constant
value in the thermodynamic limit (Fig. \ref{fig:size})
and follows the linear pressure
dependence (Equation \ref{eq:tauyield})
seen in many experimental systems.
Moreover, the shear stress is insensitive to parameters that
are not controlled in most experiments, including crystalline
alignment, coverage and sliding direction.
Even under conditions where the variations with these parameters
are largest, the changes are comparable to typical experimental
variations between different samples and
laboratories.\cite{rabinowicz65,czichos89}
This is an important test of any molecular scale model for friction.

The value of $\tau_0$ in Equation \ref{eq:tauyield} is increased
by attractive interactions between the monomers and walls or
between the two walls.
The adhesive interaction was increased both by increasing coverage
(Fig. \ref{fig:90coverage}) and by increasing $r_c$ to increase
the contribution from the attractive tail in the Lennard-Jones potential
(Fig. \ref{fig:90potential}).
Increasing the temperature (Fig. \ref{fig:f90alphatau_T})
increases the entropic repulsion and lowers $\tau_0$.

For each system there is a pressure $P_t = -\tau_0/\alpha$ below which the 
adsorbed layer acts like a liquid lubricant (Sec. \ref{sec:temp}).
When the attractive
tail of the Lennard-Jones potential is included (large $r_c$),
the value of $P_t$ for submonolayer films is negative up to quite high
temperatures.
This suggests that thicker films are needed to achieve lubricated
behavior in typical experimental systems.
However, it would be interesting to explore whether the static friction
between flat surfaces can be eliminated by the proper choice of molecule
and the use of elevated temperatures.

All of the observed trends in $\alpha$ can be understood in terms of a
simple geometrical model where monomers act like hard spheres.
Within this model, $\alpha$ only depends on the hard sphere
diameter and the wall geometry.
Parameters such as pressure, temperature, wall/monomer coupling
energy, and $r_c$ have almost no effect on the hard sphere diameter
and thus $\alpha$.
In contrast, changes in $\sigma_{\rm wf}$ or $d_{\rm nn}$ produce
direct changes in the geometry and change $\alpha$ dramatically
(Fig. \ref{fig:90potential}).
Smaller shifts in $\alpha$ are produced by
changes in coverage that alter
the ability of the hard spheres to pack into a configuration that
locks into both walls (Fig. \ref{fig:90coverage}).

Simple model potentials were used for the adsorbed molecules and
surfaces in this study, in order to allow a wide range of parameters
to be investigated.
More realistic models will have more complex interactions
and multiple atomic species.
This will complicate the geometry, but one expects that the basic
idea of an effective surface of closest approach defined by
hard-sphere repulsions will still be valid at the pressures of
interest.
This naturally produces static friction and a shear stress that
obeys Equation \ref{eq:tauyield}.
Indeed the same picture can be applied to larger particles between
bearing surfaces such as dust, sand or wear debris.

While our geometric argument for $\alpha$ bears some similarity
to earlier models based on interlocking of macroscopic\cite{dowson79}
or atomic\cite{berman98,israel00} asperities, there is
an important difference.
These previous models envisioned asperities that were rigidly
attached to their respective surfaces and would translate with them.
The mobile atoms considered here and in previous
work,\cite{muser00,muser01,he99,he01} rearrange to
create interlocking configurations for any relative position of
the surfaces.
Not only does this explain why static friction can occur between
surfaces with uncorrelated asperities, it also makes it possible
to connect static and kinetic friction.

As pointed out by Leslie\cite{dowson79} in 1804, interlocking of rigid
asperities can explain static friction, but predicts vanishing
kinetic friction.
The reason is that the energy needed to move up ramps is recovered
as the asperities move back down and the total work required to
slide the surfaces vanishes.
The situation is very different when the interlocking occurs dynamically
through motion of adsorbed molecules.
As shown in recent work,\cite{muser01,he01} at any given instant almost
all of the mobile molecules are trapped in local free energy minima.
Since lateral motion of the walls increases their energy within this
minimum they contribute to the friction.
The kinetic and static friction are closely related because the
distribution of molecules in free energy minima is nearly the same.
The difference is that when the walls are moving at fixed velocity
the minimum holding each molecule eventually becomes unstable.
It then pops forward rapidly to the nearest metastable state.
Much of its energy is dissipated in this pop because molecules
pop forward at different locations at different times in an incoherent
manner and the potential they move to varies in a random way.
It is not possible for them to move up and down over a rigid potential
energy surface as in the models of fixed asperities considered by
Leslie and earlier researchers.\cite{dowson79}


\begin{thebibliography}{10}
\bibitem{footlast}
Note that cold welding leading to static friction is observed between
some clean metal surfaces.  It may be that this is driven by
plastic deformation or that long experimental time scales allow
diffusion to reconstruct the interface into locally commensurate
grain boundaries with different orientations.
\bibitem{f3}
Note that if long-range elasticity were included in the walls one might have a
  limiting barrier at large wall areas due to activation of independent
  regions.
\bibitem{f2}
The stress is increased to this value over 10$t_{\rm LJ}$.
\end{thebibliography}
\end{document}